\documentstyle[prl,aps,amsfonts,amssymb,twocolumn,epsfig]{revtex}
\begin{document}
\draft

\newcommand{\w}{\omega}
\newcommand{\kf}{k_F}
\newcommand{\G}{{\bf G}}
\newcommand{\A}{\text{\AA}}
\newcommand{\IM}{\text{Im}}
\newcommand{\g}{\gamma}
\newcommand{\V}{V}
\newcommand{\RE}{\text{Re}}
\newcommand{\Nf}{N_0}
\wideabs{
\title{The Kondo effect in quantum dots 
at high voltage: Universality and scaling }
\author{ A. Rosch, J. Kroha and P. W\"olfle}
\address{Institut f\"ur Theorie der Kondensierten Materie, Universit\"at Karlsruhe, D-76128 Karlsruhe, Germany}
\date{\today}
\maketitle

\begin{abstract}
  We examine the properties of a dc-biased quantum dot in the
  Coulomb blockade regime. For voltages $\V$ large compared to the
  Kondo temperature $T_K$, the physics is governed by the scales $\V$
  and $\g$, where $\g \sim \V/\ln^2( \V/T_K)$ is the
  non-equilibrium decoherence rate induced by the voltage-driven
  current.  Based on scaling arguments, self-consistent perturbation
  theory and perturbative renormalization group, we argue that due to
  the large $\g$, the system can be described by renormalized
  perturbation theory for $\ln( \V/T_K) \gg 1$.  However, in certain
  variants of the Kondo problem, two-channel Kondo physics is induced
  by a large voltage $\V$.
\end{abstract}
}

In recent years, it became possible to observe the Kondo effect in
quantum dots in the Coloumb blockade regime
\cite{Gold,Cronenwett,Schmid,vanderWiel}. These systems allow to investigate,
how non-equilibrium induced by external currents and bias
voltages influences the Kondo physics. Similarly, the experimentally observed
anomalies of the energy relaxation in strongly voltage-biased
mesoscopic wires \cite{pothier.exp} have recently been shown 
\cite{wireTheo} to be caused by
scattering from magnetic impurities or two-level systems.

In equilibrium, almost all properties of the Kondo effect are
well understood, and the Kondo model together with the methods
used to solve it (e.g. renormalization group (RG), Bethe ansatz, conformal
field theory, bosonization, density matrix RG,
flow equations or slave particle techniques) 
have become one of the central paradigms in condensed
matter theory. However, in non-equilibrium 
many of the above-mentioned methods fail, and
despite the experimental and theoretical relevance
and a substantial body of theoretical work
\cite{early,Kaminski,PT,Konig,Schoeller,Schiller,Wen,Coleman,ncaPapers},
several even qualitative questions about the Kondo effect in non-equilibrium
have remained controversial.
Recently, Coleman {\it et al.} \cite{Coleman} claimed 
that the Kondo model at high voltages $V\gg T_K$
cannot be described by (renormalized) perturbation theory (PT) 
but is characterized by 
a new two-channel Kondo fixed point (see also \cite{Wen}). 
By contrast,  Kaminski et al. \cite{Kaminski} argue that the 
non-equilibrium decoherence rate $\gamma$ destroys the Kondo effect.
We will show in the following that the Kondo effect is indeed destroyed 
in the case of the usual Anderson model, but for certain variants of the
Kondo model, where the current at high bias is suppressed, the scenario proposed
in \cite{Coleman} appears to be recovered.

We model the quantum dot using the Anderson model
\begin{eqnarray}
H_A&=&H_0+
\varepsilon_d \sum_\sigma  d^\dagger_\sigma d_\sigma+
 \sum_{\alpha k \sigma}( t_\alpha c^\dagger_{\alpha k \sigma}  
d_\sigma+
h.c. ) \nonumber \\ 
&+&U n_{d\uparrow} n_{d\downarrow}
\label{HA}
\end{eqnarray}
where $H_0=\sum_{\alpha k \sigma} \varepsilon_{\alpha k}
c^\dagger_{\alpha k \sigma}c_{\alpha k \sigma}$ is the Hamiltonian of
the electrons in the left and right leads, $\alpha=L,R$, characterized
by a dc bias voltage $\V$,  $\varepsilon_{L/R k}= \varepsilon_{k} \pm \V/2$,
respectively. We will consider only symmetrical dots with tunneling matrix
elements $t_L=t_R\equiv t$.
The negative $\varepsilon_d$ with $|\varepsilon_d|\gg \Gamma =2 \pi
\Nf t ^2$, where $\Nf$ is the electron density of states
in the leads, and the large Coloumb repulsion $U\to \infty$ 
enforce the number of electrons 
$n_d=\sum_\sigma  d^\dagger_\sigma d_\sigma$
in the dot level to be approximately $1$.

In this regime,  the local degree of freedom of the
quantum dot is a spin $\vec{S}=\frac{1}{2}\sum_{\sigma,\sigma '} d_\sigma^\dagger
{\vec{\sigma}^{\phantom{\dagger}}_{\sigma\sigma '}} d_{\sigma '}$, 
where $\vec \sigma$ is the vector of
Pauli matrices, and the  low-energy properties of $H_A$ are well described
by the two-lead Kondo (or Coqblin-Schrieffer) model, 
\begin{eqnarray}
&&H_K=H_0+ V_0 \sum_\sigma 
(c^\dagger_{L0\sigma }+c^\dagger_{R0\sigma })(c_{L0\sigma }+c_{R0\sigma })
\label{HK} + \\
  &&\vec{S} \cdot  
\sum_{\sigma\sigma '}
 J_L
c_{L0\sigma}^\dagger \frac{\vec{\sigma}_{\sigma\sigma '}}{2} c_{L0\sigma '}+
J_{LR} c_{L0\sigma}^\dagger \frac{\vec{\sigma}_{\sigma\sigma '}}{2}c_{R0\sigma '}
 + (L \leftrightarrow R)
\nonumber
\end{eqnarray}
where $c_{L/R 0\sigma }=\sum_k c_{L/R k\sigma }$.  For an
Anderson model with symmetrical coupling to the leads, 
one obtains $J_L=J_R=J_{LR}=J_{RL}=4V_0 = 2t^2/\varepsilon_d \equiv J$. 
For sufficiently small $J$, the potential scattering term $V_0$ can be
neglected and, as will be seen, 
the equilibrium and the non-equilibrium
physics of the Kondo model is completely universal,
characterized by a single scale, the Kondo temperature $T_K=D
\sqrt{ \Nf J} e^{1/(2  \Nf J)}$, where $D$ is a high-energy cutoff. The
precise formula for the pre\-factor of $T_K$ depends on details of the
model. However, for $T_K, T, \V \ll D$ relevant physical quantities
like the conductance $G$ are universal, $G=G(\V/T_K,T/T_K)$ and do not
depend on details of the original Hamiltonian.

In the first part of the paper, we investigate in detail the
Anderson model in the Kondo regime at high voltages using the
so-called non-crossing approximation (NCA). In the second part we will
use the insight gained from this analysis to study a heuristic version
of poor man's scaling in non-equilibrium for a Kondo model with
$J_{LR} < J_{L/R}$.

To derive NCA, one first rewrites $H_A$ in the limit $U\to \infty$
using a so-called pseudo-fermion $f_\sigma$ and a spin-less slave
boson $b$ with $d_\sigma=b^\dagger f_\sigma$, subject to the
constraint $Q= \sum_\sigma f^\dagger_\sigma f_\sigma + b^\dagger b =1$.
The Anderson model then takes the form $H_A=H_0+\varepsilon_d b^\dagger
b+\sum_{\alpha,\sigma}( V_\alpha c^\dagger_{\alpha\sigma 0} b^\dagger
f_{\sigma}+h.c)$. 
In this language, the NCA is just the lowest-order
self-consistent PT in $t_\alpha$, 
where the constraint $Q=1$ is taken into account
exactly. While the NCA fails to describe the low-energy Fermi liquid
fixed point in the Kondo regime correctly \cite{kroha.97}, it gives 
reliable results (with errors of the order of $10\%$) in equilibrium
for temperatures down to a fraction of $T_K$.  
As a self-consistent and
conserving approximation, it also displays the correct scaling behavior and
reproduces the relevant energy scales.

\begin{figure}
\begin{center}
\epsfig{width=0.9 \linewidth,file=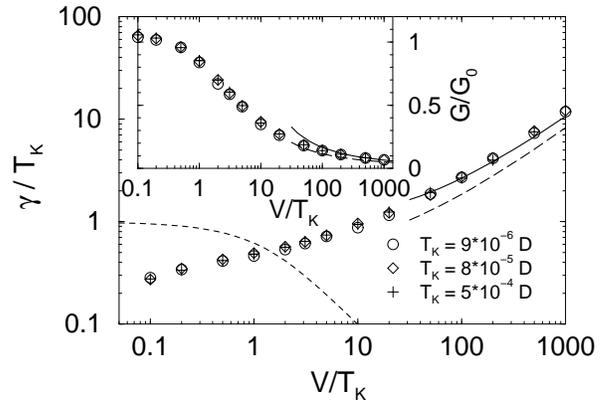}
\end{center}
\vspace*{-0.1cm}
\caption{Non-equilibrium decoherence rate
$\g=\IM \Sigma_f(0)$  calculated within NCA compared to the 
strong coupling scale $T^*=\sqrt{T_K^2+(\V/2)^2}-\V /2$ (dashed line). 
For $\g \gg T^*$ one stays in the weak coupling regime. 
The symbols correspond to 3 different values of $T_K$, $T_K / D =
 9\cdot 10^{-6}, 8\cdot 10^{-5}, 5\cdot 10^{-4} $.
Inset: conductance $G$  in units of $G_0 =2e^2/(2\pi\hbar)$.
Long-dashed and solid lines: asymptotic analytical results, 
Eqs.\ (\ref{asymGamma}), (\ref{asymG}),
in leading and next-to-leading order, respectively.\label{g.fig}}
\end{figure}

While the NCA equations in non-equilibrium have been solved by many
groups \cite{ncaPapers}, we are not aware of any careful analysis of
the relevant scales at high bias voltage, which is central for a
qualitative understanding of the non-equilibrium Kondo effect.
Generally, the NCA equations have to be solved numerically; however,
in the limit of extremely high voltage, $\ln \V/T_K \gg 1$ (but $\V \ll
D$), an analytical solution is possible: the problem is in
the weak coupling regime. 
Finite $V$ induces an inelastic spin relaxation or decoherence rate.
Since in NCA the spin density is just a convolution of the 
pseudo-fermion propagators, this rate is given by
$2 \IM \Sigma _f (0) = 2 \gamma$, with $\Sigma _f$ the pseudo-fermion 
self-energy. We start by calculating the retarded
self-energy $\Sigma^r_b(\w)$ of the boson, using the fact that 
(as shown below) the spectral function of the pseudo-fermion is a sharp
peak of width $\g \ll \V$. Throughout we consider the low temperature
limit, $T=0$, and obtain $\IM \Sigma^r_b(\w)\approx -  \pi J \Nf
|\varepsilon_d| (f_L^\g(-\w)+f_R^\g(-\w))$, where $f_{R/L}^\g$
are the Fermi functions in the left and right leads, broadened by $\g$.
The step-function in $\IM \Sigma^r_b(\w)$ leads to logarithmic
contributions to $\RE \Sigma^r_b(\w)$, cut off by $\g$ and the
band width $D$. Using relations like 
$1-2  \Nf J \ln[D/|\w|] = 2\Nf J \ln[|\w|/T_K]$ 
one obtains for the real part of the boson propagator $G_b^r(\w)$, 
for $\ln \V/T_K \gg 1$,
\begin{eqnarray}
\Nf J_{\text{eff}}^{NCA} (\w ) &\equiv& 2 \Nf t^2 \RE G_b(\w) \nonumber \\
&\approx&  \frac{1}{
 \ln\!\left( \frac{ |\w-\V/2|}{T_K}+\frac{\g}{T_K}\right)+ 
 \ln\!\left( \frac{ |\w+\V/2|}{T_K}+\frac{\g}{T_K}\right)  }
\label{Jeff}
\end{eqnarray}
This combination plays the role of an effective (frequency dependent) 
exchange coupling $J_{\text{eff}}$. 
Remarkably, the perturbative expression Eq.\ (\ref{Jeff}) 
would develop a pole close to $\w=\pm \V/2$ if $\g < 
T^*=\sqrt{T_K^2+(V/2)^2}-V/2\approx  T_K^2/V$. The breakdown scale $T^*$ of
PT has also been discussed in [4th Ref. of \cite{wireTheo}] and \cite{Coleman}.
It indicates that Eq.\ (\ref{Jeff}) is only valid for $ T^* < \g < V$.
Indeed, this criterion is fulfilled (see Fig.\ \ref{g.fig}), 
as one finds within NCA
\begin{eqnarray}
\g &\approx&  \frac{\pi}{8} \frac{\V}{\ln^2 \frac{\V}{T_K}} 
\left[1+\frac{2}{\ln  \frac{\V}{T_K}} + 
O\left( \frac{ \ln  \frac{\V}{\g}} {\ln^2  \frac{\V}{T_K}} \right)\right] \ .
\label{asymGamma}
\end{eqnarray}
For the conductance in units of the conductance quantum $G_0=2 e^2/(2
\pi \hbar)$ we obtain for $\ln \V/T_K \gg 1$
\begin{eqnarray}
\frac{G^{NCA}}{G_0} &\approx&  \frac{\pi^2}{4} \frac{1}{\ln^2 \frac{\V}{T_K}} 
\left[1+\frac{2}{\ln  \frac{\V}{T_K}} + 
O\left(\frac{\ln  \frac{\V}{\g}}{\ln^2  \frac{\V}{T_K}}\right)\right]
\label{asymG}
\end{eqnarray}

Numerical results for smaller voltages down to $V<T_K$ 
are shown in Fig.~\ref{g.fig}
and display universal behavior over the complete range of voltages 
and over several orders of magnitude in $T_K$.
Despite the fact that for high voltages, $\ln \V/T_K \gg 1$,
one stays in the weak coupling regime, the prefactors of $\g^{NCA}$ and $G^{NCA}$
are not exact, since the NCA for the Anderson model treats the potential
scattering $V_0$ and the Kondo coupling $J$ incorrectly on equal footing.
It is not difficult to obtain the correct asymptotic prefactors \cite{Kaminski}
by calculating $\g$ and $G$ in leading order PT in
$J$ for the Kondo model Eq.\ (\ref{HK}) (with $V_0=0$) and by replacing $J$
by $1/(2 \ln V/T_K)$. This corrects the leading term of
the NCA results Eqs.\ (\ref{asymGamma}), (\ref{asymG}) 
by a prefactor $3/4$. It is, however, important to
stress, that the asymptotic result in the limit $\ln V/T_K \to \infty$
is almost useless as, due to the logarithmic dependence
(Eq.\ (\ref{asymGamma})), sub-leading corrections
are very large (e.g. still $10\%$ for $\V/T_K=10^6$).

\begin{figure}
\epsfig{width=0.9 \linewidth,file=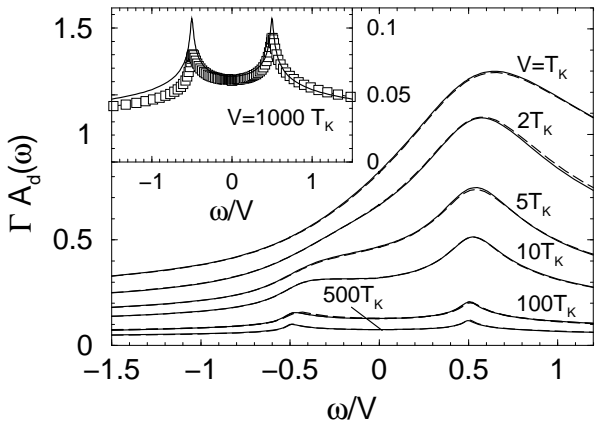}
\vspace*{0.5cm}
\caption{Spectral function $A_d^{NCA}(\w )$ 
for various voltages $V$, each calculated in NCA at two values of $T_K$
(solid and dashed lines) differing by a factor of 10. 
The NCA strongly overestimates the anisotropy of the peaks. 
Inset: asymptotic behavior of $A_d^{NCA}(\w )$
in the Kondo scaling limit. Squares: numerical
NCA result; solid line: asymptotic expression for $\ln (V/T_K) \to \infty$.
\label{a.fig}}
\end{figure}

In the limit of large $V$, the scale $\g$ influences quantities like the 
conductance, where all
electrons in an energy window $\V$ contribute, only slightly. 
The situation is different for the local spin susceptibility
on the quantum dot, $\chi  \propto 1/\g$, or the spectral function
$A_d(\omega )$ of the electron on the quantum dot. $A_d(\omega )$
calculated numerically within NCA is shown in Fig.~\ref{a.fig}. Like many
groups before, we obtain two well defined peaks at voltages $\pm V/2$.
In the limit $\text{ln}\V/T_K \to \infty$ we find approximately
$A_d^{NCA}(\w)$ $\approx$ $ (\pi^2 /\Gamma)$ $ [\Nf J_{\text{eff}}^{NCA}(\w)]^2$,
with large but universal sub-leading corrections and a non-universal,
(almost) constant potential scattering background 
of $O(\Gamma /\varepsilon _d ^2)$. 
NCA incorrectly treats potential and spin flip scattering on equal 
footing and, thus, overestimates the asymmetry 
of the peaks w.r.t.\ $\w \leftrightarrow -\w$ in the small $J$ limit. 
This can be seen from an analysis of the Schrieffer-Wolff transfor- 
\noindent
mation which shows that this asymmetry is non-universal and of
$O(\Nf V_0 )$. Since the antisymmetrical in $\omega$ contributions to
$A_d(\w)$ cancel in the integral for the conductance, non-universal
corrections to $G$ are much smaller.
Note that the logarithmic cusps of $A_d(\w)$ have an additional, small 
rounding of $O(\g )$ compared to Eq. (\ref{Jeff}), 
but for large voltage, the half width at half-maximum 
of the peaks is not given by $\g$ but by  
$\sqrt{\g\V}/2 \approx 0.3 V/\ln (V/T_K) > \g$
(see Fig.~\ref{a.fig}). 

Our analysis of the Kondo model suggests, that qualitatively different
behavior can be expected, if the non-equilibrium relaxation rate $\g$
is sufficiently small, $\g < T_K^2/V$. Since a non-zero $\g $ requires
finite current, e.g. within bare PT $\g
\propto J_{LR}^2V$, it is therefore interesting to study the Kondo
model Eq.\ (\ref{HK}) for $J_{LR}\ll J_{L}, J_{R}$, using ideas from 
perturbative RG. Such a model cannot
be derived from a simple Anderson model but may arise in more
complicated situations.  Not much is known about how
the concepts of fixed points and renormalization group can be applied
to a non-equilibrium situation (see, however, \cite{Schoeller}). The
problem is that in the presence of a finite bias voltage, many
physical quantities like the conductance are {\em not} determined by
low-energy excitations even at $T=0$, since all states with energies of
order of the applied voltage $\V$ contribute.  Therefore, a controlled
perturbative RG must probably be formulated for
the full frequency-dependent vertices in Keldish space. We will not
try to develop such a method here but propose to use a heuristic
version of poor-man's scaling adapted to the present situation. As
usual, we investigate, how coupling constants change, when the 
cutoff $\Lambda$ of the theory is modified.  As long as the cutoff is large
compared to the voltage, we expect that the usual poor-man's scaling
equations hold. For the model with $\Nf J_L=\Nf J_R=g_d$ and $\Nf
J_{LR}=g_{LR}$ and $\V \ll \Lambda$ one obtains \cite{Kaminski,Coleman}
\begin{eqnarray}
\frac{d g_d}{d \ln \Lambda} &=& - (g_d^2 + g_{LR}^2), 
\qquad \frac{d g_{LR}}{d \ln \Lambda}= - 2  g_d g_{LR} \ . \label{RG1}
\end{eqnarray}
These are the RG equations of a channel-asymmetric two-channel Kondo
model, where the even and odd channels couple to the spin with coupling
constants $g_e=g_d+g_{LR}$ and $g_o=g_d-g_{LR} \le g_e$. Note that for
the Anderson model, the odd channel decouples and $g_o=0$. 
Two parameters, $T_K$ and $\alpha$, determine the physics of the 
channel-asymmetric two-channel Kondo model,
\begin{eqnarray}
T_K&=& D e^{-1/(g_d+g_{LR})} \label{tk}, \quad
\alpha=\frac{(g_d-g_{LR})(g_d+g_{LR})}{2 g_{LR}} \ , \label{alphaDef}
\end{eqnarray}
where $T_K$ is defined by $g_e(T_K)=1$. The dimensionless number
$\alpha$ is the natural parameter to characterize the channel
anisotropy, since it is invariant under the perturbative RG flow 
Eq.\ (\ref{RG1}), i.e. $\alpha(\Lambda)=\alpha_0=const.$ for $\Lambda >
\V$. If higher orders of $g$ are included in Eqs.\ (\ref{RG1}), the
prefactor of $T_K$, Eq.\ (\ref{tk}), changes and the definition of the RG
invariant $\alpha$ has to be slightly adjusted (a dimensionless
invariant characterizing the flow will exist even in higher orders).
For the usual one-channel Kondo effect or the Anderson model Eq.\ (\ref{HA}), 
$\alpha_0 =0$, while for $\alpha_0 \to \infty$ the model is
just the well-known channel-symmetric two-channel Kondo model. We will
therefore investigate, how $\alpha$ will change for $\Lambda<V$ in order to
determine if the system flows towards a two-channel fixed point, that has
been proposed by Wen \cite{Wen} and Coleman {\it et al.}
\cite{Coleman}.  For $\V \gg T_K$, Eq.\ (\ref{RG1}) is valid down to
$\Lambda=\V$ and we obtain
\begin{eqnarray}
g_d(V)&=&\frac{1}{2} \left( \Bigl[ \ln \frac{\V}{T_K} \Bigr] ^{-1}+
         \Bigl[ \frac{1}{\alpha_0} + \ln  \frac{\V}{T_K} \Bigr] ^{-1} \right)\\
g_{LR}(V)&=&\frac{1}{2} \left( \Bigl[  \ln \frac{\V}{T_K} \Bigr] ^{-1}  -
        \Bigl[ \frac{1}{\alpha_0} + \ln  \frac{\V}{T_K} \Bigr] ^{-1}  \right) \ .
\end{eqnarray}
\noindent
For $\Lambda < \V$, the calculation of the RG flow is less obvious.
Some of the logarithmically diverging vertex corrections of 
$J$ are cut off by the voltage $V$, changing the RG flow to 
\begin{eqnarray}
\frac{d g_d}{d \ln \Lambda} &=& - g_d^2, 
\qquad \frac{d g_{LR}}{d \ln \Lambda}=0 \label{RG2} \ ,
\end{eqnarray}
in complete agreement with the analysis of Coleman {\it et al.}
\cite{Coleman}. However, all remaining logarithmic contributions are
cut off by the decoherence rate $\g$ as it is evident, 
e.g., from our analysis of NCA. 
Thus, Eqs.\ (\ref{RG2}) are only valid for $\g < \Lambda < \V$.

Since in the perturbative regime of the 
RG the bare coupling constant $\Nf J_{LR}$ is
replaced by the renormalized one, $g_{LR}$, we find
\begin{eqnarray}
\g &\sim& \V g_{LR}^2(\V) \ ,
\end{eqnarray}
which is $\V/[2 \ln (\V/T_K)]^2$ for $ \alpha_0 \ll 1/\ln (\V/T_K)$ and
\begin{figure}
\begin{center}
\epsfig{width=0.7 \linewidth,file=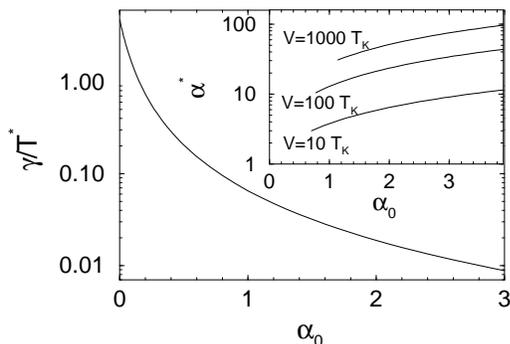}
\end{center}
\caption{$\g/T^*$ as a function of $\alpha_0$ at $\V=10 T_K$.
  Inset: $\alpha^*$ as a function of $\alpha_0$ for various voltages.
  For $\g/T^* \ll 1$, the system displays strong coupling behavior for
  $T < T^*$. $\alpha^* \gg 1$ indicates that this regime is dominated
  by two-channel physics. 
\label{rg.fig}}
\end{figure}
\noindent
$\V/[4 \alpha_0^2 \ln^4(\V/T_K)]$ for  $\alpha_0 \gg 1/\ln[\V/T_K]$. (The 
precise prefactor is irrelevant for our discussion).
If we assume for the moment that $\g$ is small, we find that 
$g_d$ flows to strong coupling  at a scale $T^*$ defined by $g_d(T^*)=1$,
\begin{eqnarray}
T^*&\approx&T_K \left(\frac{T_K}{\V}\right)^{1/[1+ 2 \alpha_0 \ln (\V/T_K)]}.
\end{eqnarray}
For $\alpha_0=0$ this scale coincides with the one introduced in
\cite{Coleman}, where  the effects of $\g$ have been
neglected.  The system will, however, only flow to strong coupling if
$\g < T^*$, while it remains in the weak coupling regime for $\g \gg
T^*$. For the usual Kondo- or Anderson model with $\alpha_0=0$, $\g$ is
always larger than $T^*$ for $\V \gg T_K$ (as $\V/T_K > \ln \V/T_K$),
and we therefore conclude in contradiction to Ref.\ \cite{Coleman} that in the 
symmetrical Kondo model there is no strong coupling regime for $V \gg T_K$.
The situation is,
however, different in the asymmetric model with $\alpha_0 \gtrsim 1/2$
(Fig.\ \ref{rg.fig}). Here the ratio
${\g}/{T^*}\approx ({\V}/{T_K})/ ({4 \alpha_0^2 \ln^4[\V/T_K]})$
is small for 
$\V \ll \V^* \approx T_K \left(4 \alpha_0^2 \ln^4\left[ 4 \alpha_0^2 
\ln^4\left[4 \alpha_0^2 \ln^4\left[ \dots \right]\right]\right]\right)$,
e.g. $V^* \approx 6 \cdot 10^{4} T_K$ for $\alpha _0 = 1$.
What is the nature of this strong coupling regime which is reached for
$T_K < \V < \V^*$ and $\alpha_0>1/2$? Insight into this question can be gained 
from a calculation of $\alpha(\Lambda =T^*)$, defined in Eq.\ (\ref{alphaDef}).
Note that in the regime $\g < \Lambda < V$, $\alpha$ is {\it not} invariant 
under the RG flow Eqs.\ (\ref{RG2}).  We obtain  
\begin{eqnarray}
\alpha^*\equiv \alpha(T^*)& \approx & \ln \frac{\V}{T_K} 
\left(1+ \alpha_0 \ln \frac{\V}{T_K}\right)
\approx \alpha_0 \ln^2 \frac{\V}{T_K} .
\end{eqnarray}
Obviously, $\alpha$ is strongly enhanced by the voltage (e.g. 
$\ln^2[10^3] \approx 50$). Since for $\alpha \to \infty$ the system maps to a
two-channel Kondo problem, we conclude that for $\alpha_0>1/2$ and $T_K < \V
<\V^*$ the system will likely be dominated by two-channel physics
over a large regime.

In this paper, we have shown that the usual Kondo model at high voltages
$T_K\ll \V \ll D$ is a weak coupling problem because relaxation processes
allowed in non-equilibrium even at $T=0$ 
destroy the building up of the Kondo effect, as their 
decoherence rate $\g \gg T^* \approx T_K^2/V$.
Nevertheless, bare perturbation theory cannot be applied and even the
leading order of renormalized perturbation theory does not give
precise results in the experimentally accessible regime due to large
sub-leading corrections.  In variants of the Kondo model with $J_{LR} <
J_{L}, J_R$, a large voltage can, however, induce a qualitatively new
behavior reminiscent of two-channel Kondo physics.

We would like to thank N. Andrei, C. Bolech, P. Coleman, C. Hooley, L.\ I.\
Glazman, W. Hofstetter, O. Parcollet and A. Zawadowski for
valuable discussions, the MPIPKS Dresden for hospitality 
during parts of this work (J.K.)
and the Emmy Noether program (A.R.) and SFB 195 
of the DFG for financial support.


\begin{references}
\vspace*{-1cm}
\bibitem{Gold} D. Goldhaber-Gordon {\it et al.}, Nature {\bf 391}, 156
  (1998).  

\bibitem{Cronenwett} S. M. Cronenwett, T. H. Oosterkamp,
  and L. P. Kouwenhoven, Science {\bf 281}, 540 (1998).

\bibitem{Schmid} J. Schmid, J. Weis, K. Eberl, and K. von Klitzing,
  Physica {\bf B258}, 182 (1998).  \bibitem{vanderWiel} W. G. van der
  Wiel {\it et al.}, Science {\bf 289}, 2105 (2000).

\bibitem{pothier.exp} H. Pothier et al.,
Phys. Rev. Lett. {\bf 79}, 3490 (1997);
F.~Pierre et al.,
{\tt cond-mat/0012038}.

\bibitem{wireTheo}
J. Kroha, Adv. Solid State Phys. {\bf 40}, 216 (2000);A.~Ka-\newline
minski, L. I. Glazman, Phys. Rev. Lett. {\bf 86}, 2400 (2001);   
G. G\"oppert, H. Grabert, {\tt cond-mat/0102150};
J. Kroha, A. Zawadowski, {\tt cond-mat/0104151}.

\bibitem{early} L. I. Glazman and M. E. Raikh, Pis'ma Zh.
Eksp. Teor. Fiz. {\bf 47}, 378 (1988) [JETP Letters {\bf 47}, 452 (1988)];
 T.K.~Ng and P. A. Lee, Phys. Rev. Lett. {\bf 61},
1768 (1988). 


\bibitem{Kaminski} A. Kaminski, Yu. V. Nazarov, and L. I. Glazman,
  Phys. Rev. Lett. {\bf 83}, 384 (1999); A. Kaminski, Yu. V. Nazarov,
  and L. I. Glazman, {\tt cond-mat/0003353}.

\bibitem{PT} 
  S. Hershfield, J. H. Davies, and J. W. Wilkins, Phys. Rev. Lett.
  {\bf 67}, 3720 (1991);
 Y. Avishai and Y. Goldin, Phys. Rev. Lett. {\bf 81}, 5394 (1998).

 
\bibitem{Konig} J. K{\"o}nig, J. Schmid, H. Sch\"oller, and G. Sch{\"o}n,
Phys. Rev. {\bf B54}, 16820 (1996). 

\bibitem{Schoeller} H. Sch\"oller and J. K{\"o}nig, Phys. Rev. Lett.
{\bf 84}, 3686 (2000); H. Sch\"oller, {\tt cond-mat/9909400}.

\bibitem{Schiller} A. Schiller and S. Hershfield, 
Phys. Rev. B {\bf 58}, 14978 (1998). 

\bibitem{Wen} X.-G. Wen, {\tt cond-mat/9812431}.

\bibitem{Coleman}
P. Coleman, C. Hooley and O. Parcollet, 
Phys. Rev. Lett. {\bf 86}, 4088 (2001).

\bibitem{ncaPapers}
 Y. Meir, N. S. Wingreen, and P. A. Lee,
Phys. Rev. Lett. {\bf 70}, 2601 (1993);
M. H. Hettler, J. Kroha and S. Hershfield,
Phys. Rev. Lett. {\bf 73}, 1967 (1994);
Phys. Rev. B {\bf 58}, 5649 (1998);  
 N. Wingreen and Y. Meir, Phys. Rev. B {\bf 49}, 11040 (1994);
P. Nordlander {\it et al.}, Phys. Rev. Lett. {\bf 83}, 808 (1999);
M. Plihal, D. C. Langreth, P. Nordlander,
Phys. Rev. B {\bf 61}, R13341 (2000).

\bibitem{kroha.97}
A theory to correct this deficiency is developed in 
J. Kroha, P. W\"olfle and T. A. Costi, Phys. Rev. Lett.\ {\bf 79}, 261 (1997);
J. Kroha and P. W\"olfle, Adv. Solid State Phys. {\bf 39}, 271 (1999).


\end{references}
\end{document}